\documentstyle[aas2pp4]{article}

\lefthead{Iyomoto et al.}
\righthead{LLAGNs in NGC~3065 and NGC~4203}

\begin{document}

\title{Low-luminosity X-ray Active Nuclei in S0 galaxies\\ NGC~3065 and NGC~4203}
\author{Naoko Iyomoto\altaffilmark{1}, Kazuo Makishima\altaffilmark{1}, Kyoko Matsushita\altaffilmark{2},\\ Yasushi Fukazawa\altaffilmark{1}, Makoto Tashiro\altaffilmark{1}, and Takaya Ohashi\altaffilmark{2}}

\altaffiltext{1}{Department of Physics, School of Science, University of Tokyo, 7-3-1 Hongo, Bunkyo-ku, Tokyo 113} 
\altaffiltext{2}{Department of Physics, School of Science, Tokyo Metropolitan University, 1-1 Minami-Ohsawa, Hachioji, Tokyo 192-03}

\begin{abstract}
We present $ASCA$ X-ray results on two S0 galaxies, NGC~3065 and NGC~4203.
In both galaxies, we detected hard X-ray emission from a point-like source at the nucleus.
Single power law model having a photon index of $\sim$1.8 well described the spectra of these sources,
while thin thermal emission which is common in S0 galaxies was not detected.
The 2--10 keV luminosities of these nuclear sources,
2.2$\times$10$^{41}$ erg s$^{-1}$ and 1.5$\times$10$^{40}$ erg s$^{-1}$,
are 1--2 orders of magnitude higher than
those expected as an assembly of low-mass X-ray binaries in these galaxies.
Our results strongly suggest that NGC~3065 and NGC~4203 host low-luminosity
active galactic nuclei (LLAGNs).
\end{abstract}


\section{Introduction}
X-ray observations with $ASCA$ (Tanaka, Inoue, \& Holt 1994) have led to the discoveries of low luminosity active galactic nuclei (LLAGNs),
with the 2--10 keV luminosity in the range 10$^{40-41}$ erg s$^{-1}$, in a fair number of nearby spiral galaxies (Makishima et al. 1994; Iyomoto et al. 1996; Ishisaki et al. 1996; Iyomoto et al. 1997).
On the contrary, case of similar LLAGNs has been scarce in elliptical or S0 galaxies. 
This is presumably due to selection effects, since
there are supporting evidences
of LLAGNs in early-type galaxies.
For one thing,
several AGNs with luminosity in the range 10$^{42-43}$erg s$^{-1}$
are known to reside in elliptical galaxies such as Cen~A (e.g. Sugizaki et al. 1997), NGC~3998 (Awaki et al. 1991) and M~87 (e.g. Schreier et al. 1982).
What is more,
radio and optical kinematics have revealed super-massive black holes in the nuclei of several nearby galaxies,
both spirals (e.g. NGC~4258; Miyoshi et al. 1995)
and ellipticals (e.g. NGC~3377 and NGC~3115; Kormendy et al. 1996).
Finally,
Ho et al. (1997a) reported that
$\sim$45 \% of S0 and elliptical galaxies
have low ionization nuclear emission region (LINER).
Since most LINER nuclei observed with ASCA have turned out to be LLAGNs,
it is expected that there are many LLAGNs in early-type galaxies.

In view of the relation between normal galaxies and AGNs, it is particularly important to search for LLAGNs in early-type galaxies, because many luminous AGNs such as quasars (Bahcall et al. 1997; Hooper et al. 1997), BL Lac objects, and radio galaxies are found in early-type galaxies.
We here report on the $ASCA$ observations of LLAGN candidates in
S0 galaxies, NGC~3065 and NGC~4203.

NGC~3065 is an S0 galaxy at a distance of 31.3 Mpc (Tully et al. 1988, in which $H_{\rm 0}$ = 75 km s$^{-1}$ Mpc$^{-1}$ is used).
No activity has been reported, except for
a moderately large luminousity in far-infrared
 (1.6 Jy at 60 $\mu$m and 1.8 Jy at 100 $\mu$m; Knapp et al. 1989).
In soft X-rays, NGC~3065 was detected with the {\it Einstein Observatory};
although the spectrum was not well determined, the 0.2--4 keV luminosity was estimated to be 9.3$\times$10$^{40}$ erg s$^{-1}$ at 31.3 Mpc (Fabbiano et al. 1992), assuming a ``Raymond-Smith'' thermal spectrum with temperature $kT$=1 keV.

NGC~4203 is an S0 galaxy at 9.7 Mpc (Tully et al. 1988, again with $H_{\rm 0}$=75 km s$^{-1}$ Mpc$^{-1}$).
It is also a far infrared source 
both at 60 $\mu$m (0.6 Jy) and 100 $\mu$m (1.9 Jy; Knapp et al. 1989).
Kim et al. (1992), using the {\it Einstein Observatory}, 
reported that NGC~4203 has a hard spectrum modeled with a bremsstrahlung having a temperature higher than 5 keV, whose 0.2--4 keV luminosity reaches 3.5$\times$10$^{40}$ erg s$^{-1}$ (Fabbiano et al. 1992).
Using $ROSAT$,
Bregman et al. (1995) detected NGC~4203 as a point-like source coincident with the nucleus,
and concluded that NGC~4203 hosts a modest AGN.
It also shows weak activities in other wavebands,
such as a weak central source at 6 cm (Fabbiano et al. 1989)
and a LINER nucleus with a broad H$_{\alpha}$ line (Ho et al. 1997b).

\section{Observations and Results}
\subsection{Observation and Data Reduction}
NGC~3065 and NGC~4203 were observed with $ASCA$ on 1995 October 22 for 76 ks and 1993 Decenber 17 for 37 ks, respectively.
In both observations, the GIS (Gas Imaging Spectrometer; Ohashi et al. 1996, Makishima et al. 1996)
was operated in PH normal mode, and the SIS (Solid-State Imaging Spectrometer; Bruke et al. 1991) was operated in 2-CCD mode.
Both targets were clearly detected with $ASCA$.
We corrected the SIS data of NGC~3065 for residual dark distribution (RDD) 
using the frame-mode data obtained on 1995 November 20 and 21.
The mean counting rates of NGC~3065 were 0.06 and 0.08 c s$^{-1}$ with the GIS and SIS, respectively. Those of NGC~4203 were 0.02 c s$^{-1}$ with the GIS, and 0.04 c s$^{-1}$ with the SIS.

We selected the data which satisfied the following criteria.
First, the geomagnetic cut-off rigidity should be larger than 7 GeV c$^{-1}$ for the GIS data and 6 GeV c$^{-1}$ for the SIS data.
Second, the source elevation angle above Earth's limb during periods of dark Earth should be larger than 5$^{\circ}$ for the GIS and 10$^{\circ}$ for the SIS.
Third, for SIS only, the source elevation angle should be larger than 25$^{\circ}$ for sunlit earth.

\subsection{X-ray images}
In Figure 1, we show contour maps of the SIS images of NGC~3065 and NGC~4203.
In the NGC~3065 image, the peak of X-ray emission coincides in position with the optical nucleus (10$^{\rm h}$01$^{\rm m}$53$^{\rm s}$, $+$72$^{\circ}$10$'$13$''$) within 0$'$.3,
which is within the typical astrometric accuracy of $ASCA$ ($\sim1'$).
In the NGC~4203 image, we see two point-like sources,
in agreement with the $ROSAT$ image obtained in 1991 (Bregman et al. 1995).
The north and south brightness centroids coincide in position with the optical nucleus of NGC~4203 (12$^{\rm h}$15$^{\rm m}$21$^{\rm s}$, $+$33$^{\circ}$10$'$22$''$) within 0$'$.2 and with a blue object TON~1480 (or CSO~400) at (12$^{\rm h}$15$^{\rm m}$09$^{\rm s}$, $+$33$^{\circ}$09$'$56$''$; Pesch \& Sanduleak 1988) within 0$'$.3, respectively.

Let us next examine the angular extent of the X-ray emission.
We projected the NGC~3065 image onto y axis of the SIS~0 detector,
while the NGC~4203 image onto a dashed line in Figure 1b.
We compared the projected SIS~0 brightness profiles
with those of point spread function (PSF) in 0.6--7.0 keV 
convolved through Gaussian profiles of various widths.
In analyzing the NGC~4203 nucleus or the south source,
we fixed the Gaussian width of the other to 0$'$.0.
All these sources were confirmed to be point-like, because their profiles were fitted successfully with Gaussians of zero width (Figure 2).
The 90\% confidence upper limits on the source angular extent of NGC~3065, the NGC~4203 nucleus, and the south source are 0$'$.1 (or 0.9 kpc at a distance of 31.3 Mpc), 0$'$.2 (or 0.6 kpc at a distance of 9.7 Mpc), and 0$'$.2, respectively.

\subsection{X-ray spectra}
\subsubsection{NGC~3065}
We made the GIS and SIS spectra of NGC~3065, integrating the photons around the nucleus with radius of 4$'$ and of 3$'$, respectively.
We made background spectra
using the blank sky data at the same positions on the detector plane as the on-source spectra, 
and subtracted them from the on-source spectra.
The background count rate was $\sim$13\% of the total count rate.
We then fitted the GIS (GIS~2+GIS~3) and the SIS (SIS~0+SIS~1) spectra separately, with an absorbed power-law model.
The photon index, absorption, and flux estimated from the two instruments are consistent with each other within 90 \% error regions.
Accordingly, we next fitted the GIS and SIS spectra jointly
to find that the data were well described by a photon index of 1.8.
We summarize the fit parameters in Table 1,
and present the X-ray spectra of NGC~3065 in Figure 3a
together with the best-fit power-law model obtained from the joint fit.
At a distance of 31.3 Mpc, the 2--10 keV luminosity becomes (2.2$\pm$0.1)$\times$10$^{41}$ erg s$^{-1}$.

Alternatively, a bremsstrahlung model of temperature $\sim$6 keV, together with the Galactic absorption,
also well accounted for the GIS, SIS, and the GIS+SIS spectra of NGC~3065.

In order to study the Fe-K emission line feature,
we added a narrow gaussian component,
having a fixed center energy at 6.4 keV or 6.7 keV, to the power law model.
The 90\% upper limit of the equivalent width of the 6.4 keV (neutral) line turned out to be 0.2 keV, and that of 6.7 keV (helium-like) line was 0.3 keV.

\subsubsection{NGC~4203}
Since the separation between the NGC~4203 nucleus 
and the south source is only 2$'$, 
we accumulated the GIS and SIS events around each of these sources 
using a small integration radius of 1$'$. 
Since half power diameter of the $ASCA$ XRT is $\sim 3'$,
some signal mixing occurs between the two integration regions.
We consider this effect later.
The ratio of the background count to the 
total count of the nuclear region was $\sim$4 \%,
and that of the south-source region was $\sim$5 \%.

Figure 3b shows X-ray spectra obtained from the nuclear region.
These spectra are well described by 
either a power-law model of photon index $\Gamma \sim$1.8,
or a bremsstrahlung model of temperature $kT \sim$6 keV.
The spectral parameters estimated separately with the two instruments 
are consistent with each other within 90 \% error regions.
Accordingly, we jointly fitted the GIS and SIS spectra 
keeping all parameters common,
except the model normalization in order to cope with 
the difference in spatial resolution between the two instruments.
The best-fit power-law model obtained from the joint fit 
is overlaid on the spectra of Figure 3b,
and its parameters are summarized in Table 1.
The Fe-K line was not statistically significant:
in the same manner as NGC~3065, 
the upper limits on the Fe-K equivalent width for neutral and helium-like line 
were estimated to be 0.2 keV and 0.3 keV, respectively.

The spectrum obtained from the south-source region 
also shows featureless spectra (Figure 3c). 
Either a power-law model or a bremsstrahlung model 
well accounts for the GIS and SIS spectra. 
The best-fit power-law model determined jointly 
with the GIS and SIS is overlaid on Figure 3c,
and the derived spectral parameters are presented in Table 1.

We now correct the above results for the signal mixing 
between the two integration regions.
Fortunately, the spectral shape parameters derived from these two regions 
are very similar to each other, as can be seen in Table 1.
Therefore the spectral shapes of the two sources 
are not affected by the signal mixing, 
and we only need to make corrections to their fluxes.
Utilizing the point spread functions (PSFs) of the XRT+GIS and the XRT+SIS,
we estimate that $\sim$25 \% and $\sim$20 \% of the photons 
detected with the GIS and SIS in the nucleus region 
are coming from the south source, respectively.
Conversely, $\sim$35 \% (GIS) and $\sim$30 \% (SIS) of the photons
detected in the south-source region are in reality from the nucleus.
By sorting out these effects,
we have calculated the intrinsic fluxes of the two sources as shown in Table 1.
These values are weighted means of the two instruments,
since they gave consistent results.
Thus, the 2--10 keV luminosity of the nuclear source becomes 
$(1.5^{+0.3}_{-0.2})\times 10^{40}$ erg s$^{-1}$ at a distance of 9.7 Mpc.

These two sources were studied previously with $ROSAT$ (Bregman et al. 1995).
The absorptions estimated with $ROSAT$ 
of 2.5$\times$10$^{20}$cm$^{-2}$ (nucleus) and 3.4$\times$10$^{20}$cm$^{-2}$ 
(south source) are within 90 \% error regions of those with $ASCA$, 
while the photon indices of 2.24 (nucleus) and 2.84 (south source) are 
somewhat softer than the $ASCA$ results, although Bregman et al. (1995) 
did not mention errors associated with the photon indices.

\subsection{Time variability}
We made 0.7--8 keV GIS light curves of the three sources detected in the two fields (Figure 4).
The photon integration regions are 3$'$ around NGC~3065 and 1$'$ around the two sources in NGC~4203.
The ratio of the background to the signal is $\sim$10 \% (NGC~3065) or $\sim$5 \% (the NGC~4203 nucleus and the south source).
We did not use the SIS data,
since the sources are placed near (1$'$) to the dead region between two chips in the SIS detectors
and hence the SIS count rates were subject to the spacecraft attitude jittering,
although it has a minor effects on the spectral analysis.
As the Figure indicates, no significant short-term variability was detected above an upper limit of 50 \% (peak-to-peak) during 
each observation.
If we fit the data assuming a constant count rate,
$\chi^2$/d.o.f. of the GIS light curves become 70/75 (NGC~3065), 68/63 (the NGC~4203 nucleus), and 62/63 (the south source).

In order to investigate long-term variability,
we extrapolated the power-law models of the $ASCA$ spectra toward lower energies, 
considering errors of photon index and intrinsic absorption,
and compared the calculated fluxes with those obtained with the
{\it Einstein Observatory} and $ROSAT$.
As shown in Table 2,
the 0.2--4.0 keV flux of NGC~3065 predicted by $ASCA$
is thus twice higher than that measured with the {\it Einstein Observatory} in 1980.
However this difference is likely to be a result of the inadequate modeling of the {\it Einstein} flux; 
it was estimated assuming 1 keV thin thermal emission,
which is much softer than those measured with $ASCA$ 
and give rise to an underestimation of the flux by a factor $\sim$1.8.
Besides, the {\it Einstein} flux of NGC~3065 has a large error
because the exposure time was short (0.9 ks) and total counts were very small (25.2).

The NGC~4203 nucleus and the south source were not resolved with {\it Einstein} due to a short exposure time (1.5 ks).
Therefore, we compared the summed flux of these sources predicted by $ASCA$ to the flux of whole galaxy measured with {\it Einstein}.
The latter was obtained assuming an adequate model,
namely, a Raymond-Smith model with $kT$ greater than 5 keV and solar abundance.
Considering relatively small total counts (189.97) with the {\it Einstein Observatory},
the fluxes measured with $ASCA$ and the {\it Einstein Observatory}
agree with each other within model uncertainties.
The fluxes of the NGC~4203 nucleus and the south source observed with $ROSAT$ and $ASCA$
are also consistent with each other within model uncertainties.
In short, there is no clear evidence of long-term variability in
any of the three sources.

\section{Discussion}
The spectra of NGC~3065 and the NGC~4203 nucleus
have been well modeled with a power-law having a Seyfert-like photon index of 1.8.
Besides, both sources are pointlike within the limitation of the instrumental PSF,
and their positions are consistent with those of the optical nuclei.
Therefore these X-ray sources are likely to be faint active nuclei. 

We can not however rule out
another possibility that these sources are 
attributed to unresolved integrated emission from
a bunch of low mass X-ray binaries (LMXBs)
which underlies the X-ray spectra of all galaxies (Canizares et al. 1987; Matsushita et al. 1994).
Actually, a Bremsstrahlung model,
which is an approximation to the X-ray spectra from LMXBs (Makishima et al. 1989),
describes the spectra of these two objects
as adequately as the power-law model.
The constraints
on the angular extent is relatively loose.
Moreover we detected no time variability.

To distinguish between the LLAGN and LMXB emissions,
we may utilize the fact that the luminosity ot the latter component is well correlated with the optical luminosity of the host galaxy (Canizares et al. 1987; Matsushita et al. 1994).
For this purpose,
in Figure 5 we plotted 
the 2--10 keV luminosity $L_{\rm X}$ of the two sources at the nuclei against 
the blue band luminosity $L_{\rm B}$ of the host galaxies
in comparison with those of normal early-type galaxies (Matsushita et al. 1994) and LLAGNs in spiral galaxies.
Spectra of the latter two groups usually consist of two components; a thin-thermal emission from inter-staller matter and a harder emission from LMXB and/or AGN.
Therefore we used 2--10 keV luminosities of the hard component $L_{\rm X}^{\rm hard}$ only in the comparison,
utilizing spectral studies with ASCA.
Thus, the $L_{\rm X}$/$L_{\rm B}$ ratios of NGC~3065 and NGC~4203 are two orders of magnitude higher than those of normal galaxies,
and even exceed those of spiral-hosted LLAGNs.
We may, therefore, reasonably attribute these X-ray sources to the LLAGN emission,
wherein the LMXB contribution is negligible.

We thus conclude that the two S0 galaxies, NGC~3065 and NGC~4203, both host LLAGNs. This conclusion on NGC~4203 confirms the previous suggestion by Bregman et al. (1995),
where that on NGC~3065 is an unexpected new finding. 
These results encourages us to speculate that LLAGNs are abundant not only in late-type galaxies but also in early-type ones.

The X-ray spectra of these LLAGNs require small or no intrinsic absorption; consequently, they are Type I objects.
The insignificance of fluorescent Fe-K lines are also consistent with the results of the Seyfert I's (Nandra et al. 1997).
The photon index of 1.8 also suggests that the nuclear emission is seen directly,
because if the reprocessed emission were dominant, the spectrum would become harder than observed.
NGC~1097 (Iyomoto et al. 1996) and M~81 (Ishisaki et al. 1996) are similar examples of Type I LLAGNs in late-type galaxies.

The nuclei of NGC~3065 and NGC~4203 show no significant short-term X-ray variability.
Similar tendency is reported in several late-type LLAGNs,
and interpreted as a possible evidence of larger black hole mass.
Thus, the X-ray properties of these two LLAGNs in s0 galaxies are not different from those in late-type galaxies,
except that their $L_{\rm X}$/$L_{\rm B}$ ratios are somewhat higher.

NGC~3065 and NGC~4203 are moderately luminous in far infrared, 
and ratios of the 0.5--4.0 keV flux to 60 $\mu$m flux
become 5.5$\times$10$^{-2}$ and 1.4$\times$10$^{-2}$, respectively.
These lie on the X-ray to 60 $\mu$m correlation among active galaxies reported in Green et al. (1992).

The south source coincides in position with a blue object TON~1480 (or CSO~400) at an apparent blue magnitude of 17 (Pesch \& Sanduleak 1988).
It is also detected at 4.85 GHz with a flux of 36$\pm$7 mJy (Becker et al. 1991; Gregory \& Condon 1991).
These properties, together with the point-like image, hard spectrum, and a 2--10 keV flux of (1.0$^{+0.3}_{-0.2}$)$\times10^{-12}$ erg s$^{-1}$ cm$^{-2}$ obtained with $ASCA$,
are consistent with an interpretation 
that it is a background AGN.

Finally, we searched for 
thin-thermal emission
from the two galaxies,
by adding a Raymond-Smith model with fixed temperature (1 keV) and abundance (0.5 solar) to the power-law model.
We obtained upper limits only; 6.4$\times10^{40}$ erg s$^{-1}$ and 1.4$\times10^{39}$ erg s$^{-1}$ in 0.5--4 keV for NGC~3065 and NGC~4203, respectively.
These upper limits are reasonable for plasma emissions from inter-stellar matter in S0 galaxies (Matsushita et al. 1994), 
implying that the thermal emission is completely masked by the strong AGN emission.

\figcaption{X-ray contour maps of (a) NGC~3065 and (b) NGC~4203, taken with the SIS (S0 + S1) in the 0.7--7 keV band. Sky coordinates are J2000. \label{fig1}}

\figcaption{Comparison between the 0.7--8 keV projected SIS~0 profile
and the instrumental point-spread function (PSF).
(a) Results on NGC~3065.
Crosses represent the SIS-0 data projected onto y-axis of the detector,
while the histogram shows the PSF plus the background.
(b) The same as (a), but for NGC~4203.
The SIS-0 data are projected onto the dashed line in figure 1b.
The left and right peaks represent the south source and the nucleus, respectively. \label{fig2}}

\figcaption{(a) The top panel shows the GIS (S2+S3; thick line) and SIS (S0+S1; thin line) spectra of NGC~3065 with a power-law model determined jointly by the GIS + SIS. 
They are shown after background subtraction, but without removing the instrumental response.
The middle and bottom panels show the GIS residuals and SIS residuals, respectively. (b) The same as (a), but for the NGC~4203 nucleus. (c) The same as (a), but for the source placed at 2$'$ south of NGC~4203. \label{fig3}}

\figcaption{Background-inclusive 0.7--8 keV light curves from GIS (S2+S3) of
(a) NGC~3065, (b) the NGC~4203 nucleus, and (c) the south source.
Each bin consists of 1000 s exposures.
The straight lines indicate the average count rates.
Background rate is $\sim$4$\times$10$^{-3}$ c s$^{-1}$ for (a), 
and $\sim$8$\times$10$^{-4}$ c s$^{-1}$ for (b) and (c). \label{fig4}}

\figcaption{Comparison between the 2--10 keV luminosity $L_{\rm X}$ against the blue band luminosity $L_{\rm B}$ of
NGC~3065 and NGC~4203 (filled circles).
As to NGC~4203, $L_{\rm X}$ refers to the nuclear source only,
while $L_{\rm B}$to the entire galaxy.
For comparison, normal early-type galaxies (open circles; Matsushita et al. 1994) and LLAGNs (crosses; Makishima et al. 1994; Iyomoto et al. 1996; Ishisaki et al. 1996; Iyomoto et al. 1997) are also plotted. \label{fig5}}

\begin{table*}
\caption{Best fit parameters of NGC~3065, the NGC~4203 nucleus and the south source. \label{tbl-1}}
\begin{center}
\begin{tabular}{lllllll} \tableline \tableline
&\multicolumn{2}{c}{NGC~3065}
&\multicolumn{2}{c}{NGC~4203 nucleus}
&\multicolumn{2}{c}{south source}\\
&Power-law&Brems.&Power-law&Brems.&Power-law&Brems.\\ \tableline
$\Gamma$ or $kT$ [keV]
				&$1.80\pm0.05$&$6.1_{-0.4}^{+0.5}$
				&$1.85^{+0.10}_{-0.09}$&$ 5.7_{-0.7}^{+0.9}$
				&$1.90_{-0.07}^{+0.11}$&$ 4.4_{-0.5}^{+0.6}$\\
2--10 keV flux \tablenotemark{a}
				&$1.9\pm0.1$&$1.8\pm0.1$ 
				&$1.4\pm0.2$ \tablenotemark{b}&$1.3\pm0.1$ \tablenotemark{b}
				&$1.0_{-0.2}^{+0.3}$ \tablenotemark{b}&$0.8\pm0.2$ \tablenotemark{b}\\
Excess absorption \tablenotemark{c}
				&$4\pm3$&$0\pm0$
				&$5_{-4}^{+5}$&$0_{-0}^{+1}$
				&$0_{-0}^{+5}$&$0_{-0}^{+1}$\\ 
$\chi^2$/d.o.f.			&201/198	&232/198
				&94/145		&105/145
				&118/154	&144/154\\ \tableline
\end{tabular} 
\end{center}

\tablecomments{The GIS and the SIS spectra are fitted jointly. 
      Errors refer to single-parameter 90\% confidence limits.
}

\tablenotetext{a}{In unit of 10$^{-12}$ erg s$^{-1}$ cm$^{-2}$.}
\tablenotetext{b}{Averaged values of the GIS and SIS.
Shown after subtraction of the contamination due to the flux mixing.}
\tablenotetext{c}{In unit of 10$^{20}$ cm$^{-2}$.
	The Galactic $N_{\rm H}$ is fixed at 3.4 and 1.2
	$\times$10$^{20}$ cm$^{-2}$ for NGC~3065 and NGC~4203, respectively.}

\end{table*}

\begin{table*}
\caption{Soft X-ray fluxes of NGC~3065 and NGC~4203. \label{tbl-2}}
\begin{center}
\begin{tabular}{lllllll}\tableline\tableline
&\multicolumn{2}{c}{NGC~3065}&\multicolumn{4}{c}{NGC~4203}\\ \cline{4-7}
				&	&			&	&whole		&nucleus	&south source\\ \tableline
				&epoch	&0.2--4.0 keV		&epoch	&0.2--4.0 keV	&0.1--2.5 keV	&0.1--2.5 keV\\
Einstein			&1979	&0.79			&1980	&3.1		&---		&---	\\
{\it ROSAT}			&---	&---			&1991	&---	&2.2		&1.3	\\
{\it ASCA}\tablenotemark{a}	&1995	&1.9$^{+0.2}_{-0.1}$	&1993	&3.0$^{+0.8}_{-0.6}$ \tablenotemark{b}	&1.2$^{+0.5}_{-0.2}$ \tablenotemark{b}	&1.1$^{+0.3}_{-0.4}$ \tablenotemark{b}\\ \tableline
\end{tabular}
\end{center}

\tablecomments{Shown without removing Galactic and intrinsic absorptions, in unit of 10$^{-12}$ erg s$^{-1}$ cm$^{-2}$.}

\tablenotetext{a}{Estimated from the joint fit to the GIS and SIS spectra.}
\tablenotetext{b}{Averaged values of the GIS and SIS.
Shown after subtraction of the contamination due to the flux mixing.}

\end{table*}

\begin{references}
\reference{awaki1991} Awaki, H., Koyama, K., Kunieda, H., Takano, S., Tawara, Y., and Ohashi, T. 1991, ApJ, 366, 88  
\reference{bahcall1997} Bahcall, J.N., Kirhakos, S., Saxe, D.H., and Schneider, D.P. 1997, ApJ, 479, 642 
\reference{becker1991} Becker, R.H., White, R.L., and Edwards, A.L. 1991, ApJS, 75, 1 
\reference{bregman1995} Bregman, J.N., Hogg, D.E., and Roberts, M.S. 1995, ApJ, 441, 561
\reference{burke1991} Burke, B.E., Mountain, R.W., Harrioson, D.C., Bautz, M.W., Doty, J.P., Ricker, G.R., and Daniels, P.J. 1991, IEEE Trans. ED-38 p1069 
\reference{cani1987} Canizares, C.R., Fabbiano, G., and Trinchieri, G. 1987, ApJ, 312, 503  
\reference{fabbiano1992} Fabbiano, G., Kim, D.W., and Trinchieri, G. 1992, ApJS, 80, 531 
\reference{fabbiano1989} Fabbiano, G., Gioia, I.M., and Trinchieri, G. 1989, ApJ, 347, 127
\reference{green1992} Green, P.J., Anderson, S.F., and Ward, M.J. 1992, MNRAS, 254, 30 
\reference{gregory1991} Gregory, P.C., and Condon, J.J. 1991, ApJS, 75, 1011 
\reference{ho1997a} Ho, L.C., Filippenko, A.V., Sargent, and W.L.W. 1997a, ApJ, 487, 568 
\reference{ho1997b} Ho, L.C., Filippenko, A.V., Sargent, W.L.W., and Peng, C.Y. 1997b, ApJS, 112, 391
\reference{hooper1997} Hooper, E.J., Impey, C.D., and Foltz, C.B. 1997, ApJ, 480, 95 
\reference{ishisaki1996} Ishisaki, Y., Makishima, K., Inoue, H., Iyomoto, N., Kohmura, Y., Mitsuda, K., Mushotzky, R.F., Petre, R. et al. 1996, PASJ, 48, 237 
\reference{iyomoto1997} Iyomoto, N., Makishima, K., Fukazawa, Y., Tashiro, M., \& Ishisaki, Y. 1997, PASJ, 49, 425
\reference{iyomoto1996} Iyomoto, N., Makishima, K., Fukazawa, Y., Tashiro, M., Ishisaki, Y., Nakai, N., and Taniguchi, Y. 1996, PASJ, 48, 231
\reference{kim1992} Kim, D.W., Fabbiano, G., and Trinchieri, G. 1992, ApJ, 393, 134 
\reference{knapp1989} Knapp, G.R., Guhathakurta, P., Kim, D.W., and Jura, M. 1989, ApJS, 70, 329 
\reference{kormendy1995} Kormendy, J. 1995, ARAandA, 33, 581
\reference{makishima1989} Makishima, K., Ohashi, T., Hayashida, K., Inoue, H., Koyama, K., Takano, S., Tanaka, Y., Yoshida, A., et al. 1989, PASJ, 41, 697 
\reference{makishima1994} Makishima, K., Fujimoto, R., Ishisaki, Y., Kii, T., Loewenstein, M., Mushotzky, R., Serlemitsos, P., Sonobe, T. et al. 1994, PASJ, 46, L77 
\reference{makishima1996} Makishima, K., Tashiro, M., Ebisawa, K., Ezawa, H., Fukazawa, Y., Gunji, S., Hirayama, M., Idesawa, E. et al. 1996, PASJ, 48, 171
\reference{matsushita1994} Matsushita, K., Makishima, K., Awaki, H., Canizares, C.R., Fabian, A.C., Fukazawa, Y., Loewenstein, M., Matsumoto, H. et al. 1994, ApJ, 436, L41	
\reference{miyoshi1995} Miyoshi, M., Moran, J., Herrnstein, J., Greenhill, L., Nakai, N., Diamond, P., \& Inoue, M. 1995, Nature, 373, 127
\reference{nandra} Nandra, K., George, I.M., Mushotzky, R.F., Turner, T.J., and Yaqoob, T. 1997, ApJ, 477, 602 
\reference{ohashi1996} Ohashi, T., Ebisawa, K., Fukazawa, Y., Hiyoshi, K., Horii, M., Ikebe, Y., Ikeda, H., Inoue, H. et al. 1996, PASJ, 48, 157 
\reference{pesch1988} Pesch, P., and Sanduleak, N. 1988, ApJS, 66, 297 
\reference{schreier1982} Schreier, E.J., Gorenstein, P., and Feigelson, E.D. 1982, ApJ, 261, 42 
\reference{sugizaki1997} Sugizaki, M., Inoue, H., Sonobe, T., Takahashi, T., and Yamamoto, Y. 1997, PASJ, 49, 59 
\reference{tanaka1994} Tanaka, Y., Inoue, H., \& Holt, S. 1994, PASJ, 46, L37
\reference{tully1988} Tully, R.B. 1988, Nearby galaxies catalog (Cambridge University Press, New York) 
\end{references}
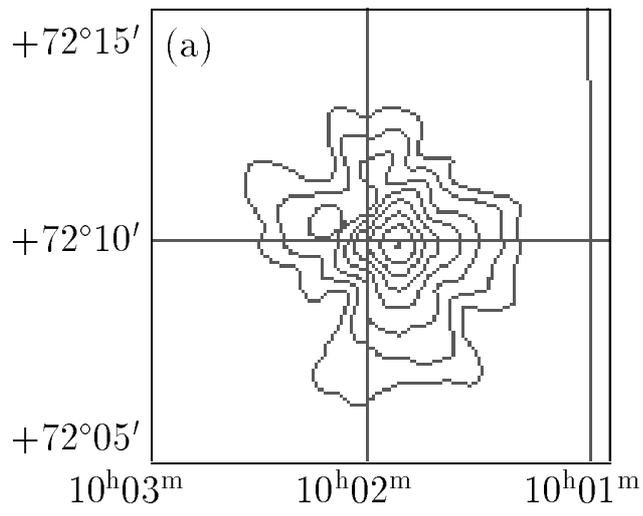
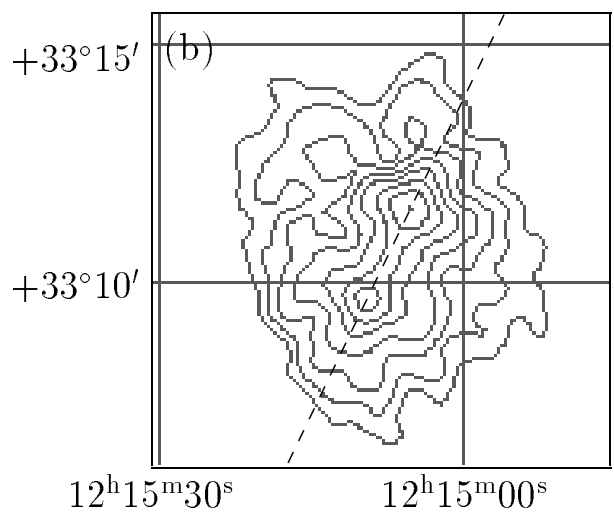
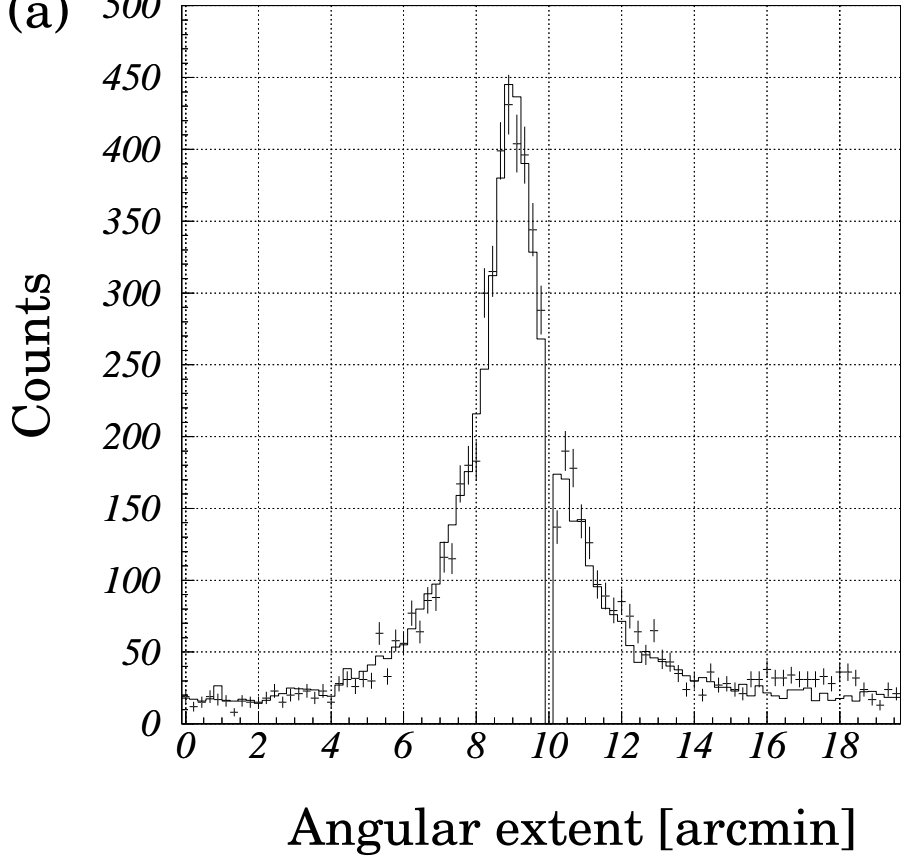
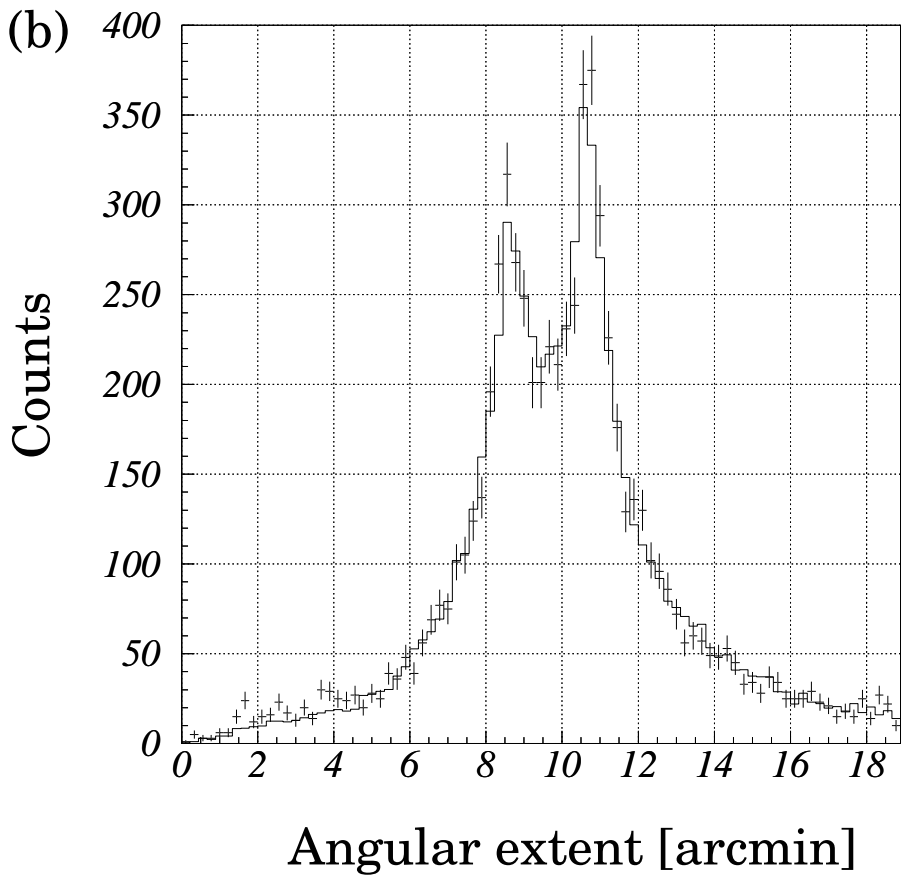
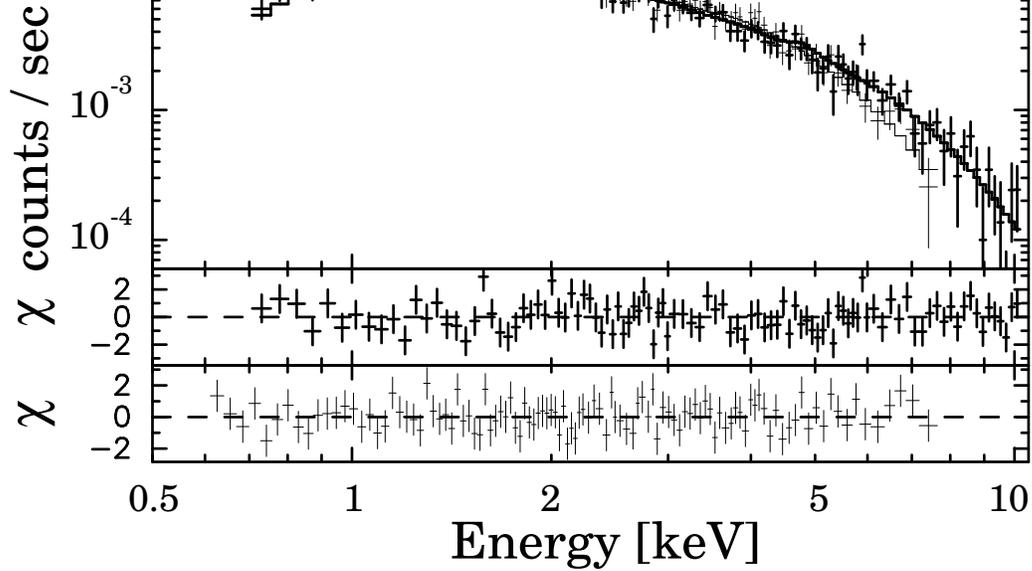
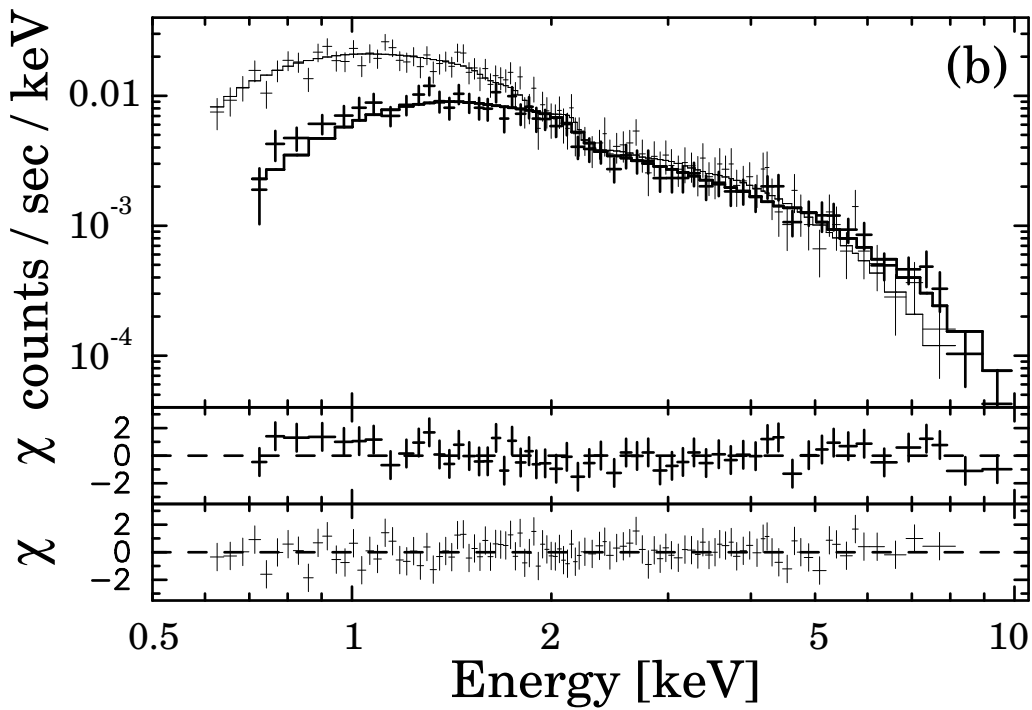
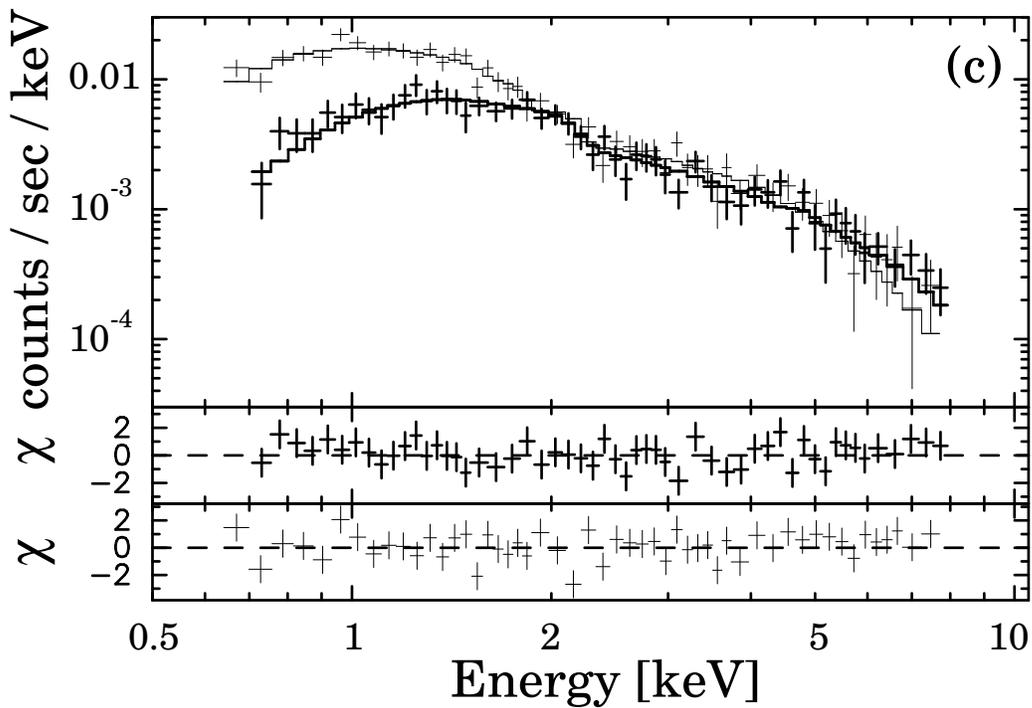
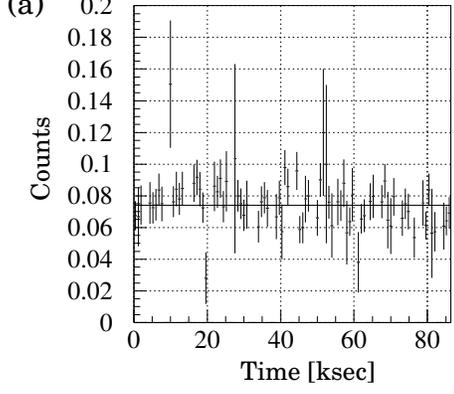
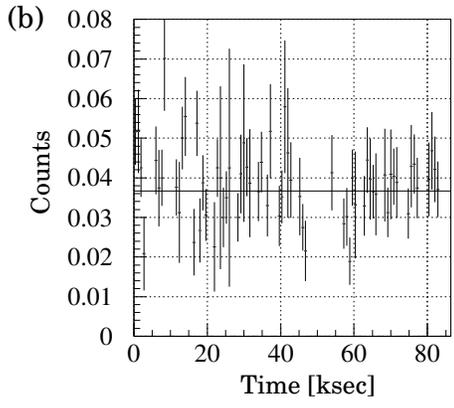
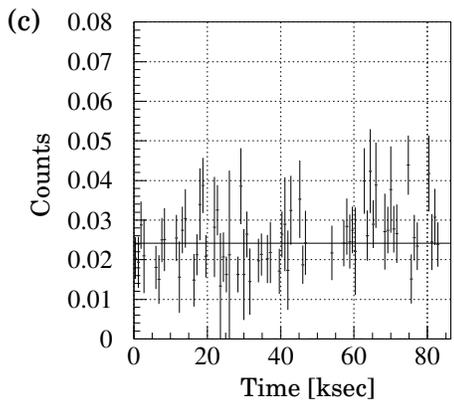
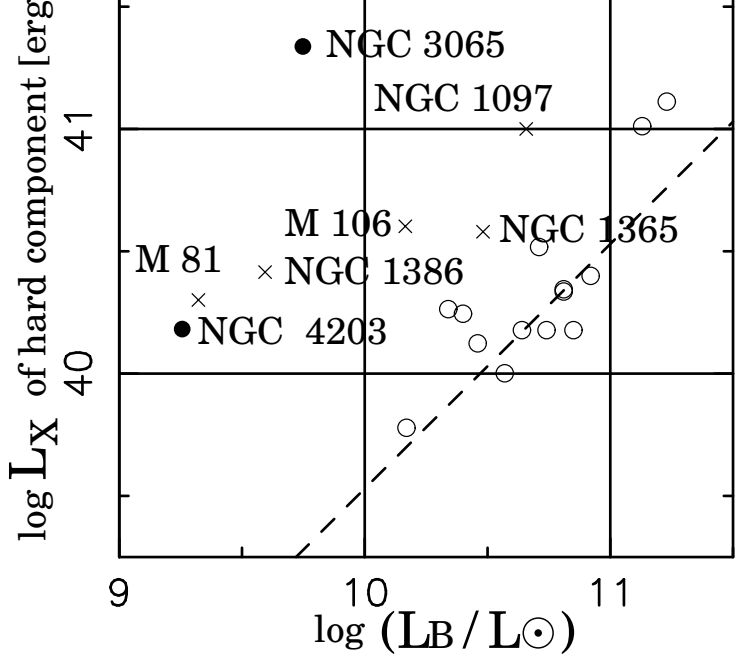
\end{document}